\documentclass[twocolumn,prl,showpacs]{revtex4}
\usepackage{graphicx}
\usepackage{dcolumn}
\usepackage{bm}
\usepackage{amsmath}
\usepackage{times}

\begin{document}

\preprint{Odd-frequency proximity}
\title{Theory of the proximity effect in junctions with unconventional superconductors}
\author{ Y. Tanaka$^{1,2}$ and A. A. Golubov$^{3}$ }
\affiliation{$^1$Department of Applied Physics, Nagoya University, Nagoya, 464-8603,
Japan \\
$^2$ CREST Japan Science and Technology Cooperation (JST) 464-8603 Japan \\
$^3$ Faculty of Science and Technology, University of Twente, The
Netherlands \\
}
\date{\today}
\begin{abstract}
We present a general theory of the proximity effect in junctions between diffusive normal metals (DN) and superconductors. Various possible symmetry classes in a superconductor are considered: even-frequency spin-singlet even-parity (ESE) state, even-frequency spin-triplet odd-parity (ETO) state, odd-frequency spin-triplet even-parity (OTE) state and odd-frequency spin-singlet odd-parity (OSO) state. It is shown that the pair amplitude in a DN belongs respectively to an ESE, OTE, OTE and ESE pairing state since only the even-parity s-wave pairing is possible due to the impurity scattering. 
\end{abstract}

\maketitle



%
%

%



It is well established that superconductivity is realized due to the
formation of Cooper pairs consisting of two electrons. In accordance with
the Pauli principle, it is customary to distinguish spin-singlet even-parity
and spin-triplet odd-parity pairing states in superconductors, where odd
(even) refer to the orbital part of the pair wave function. For example, $s$
-wave and $d$-wave pairing states belong to the former case while $p$-wave
state belongs to the latter one \cite{Sigrist}. In both cases, the pair
amplitude is an even function of energy. However, the so-called
odd-frequency pairing states when the pair amplitude is an odd function of
energy can also exist. Then, the spin-singlet odd-parity and the
spin-triplet even-parity pairing states are possible.

The possibility of realizing the odd-frequency pairing state was first
proposed by Berezinskii in the context of $^{3}$He, where the odd-frequency
spin-triplet hypothetical
pairing was  discussed \cite{Berezinskii}.
The possibility of the odd-frequency
superconductivity was then discussed in the context of various mechanisms of
superconductivity involving strong correlations \cite{Balatsky,Fuseya}.
There are several experimental evidences \cite{Bulk}
which are consistent with
the realization of the odd-frequency
bulk superconducting state in Ce compounds \cite{Bulk,Fuseya}.
In more accessible systems
(ferromagnet/superconductor heterostructures with inhomogeneous
magnetization)
the odd-frequency pairing state was first proposed in
Ref. \onlinecite{Efetov1} and then
various aspects of this state were intensively
studied \cite{Efetov2}.
At the same time, the very important issue of the manifestation of
the odd-frequency pairing in proximity systems without magnetic
ordering received no attention yet. This question is addressed in
the present Letter.

Coherent charge transport in structures involving diffusive normal metals
(DN) and superconductors (S) was extensively studied during the past decade both experimentally and theoretically. However, almost all previous work was
restricted to junctions based on conventional $s$-wave superconductors \cite%
{Proximity}. 
%
Recently, new theoretical approach to study charge transport in junctions
based on $p$-wave and $d$-wave superconductors was developed and applied to
the even-frequency pairing state \cite{Proximityd,Proximityp}. It is known
that in the anisotropic paring state, due to the sign change of the pair
potential on the Fermi surface, a so-called midgap Andreev resonant state
(MARS) is formed at the interface \cite{TK95,Buch}. As was found in \cite%
{Proximityd,Proximityp}, MARS competes with the proximity effect in contacts
with spin-singlet superconductors, while it coexists with the proximity
effect in junctions with spin-triplet superconductors. In the latter case,
it was predicted that the induced pair amplitude in the DN has a peculiar
energy dependence and the resulting local density of states (LDOS) has a
zero energy peak (ZEP) \cite{Proximityp}.
However, the relation of this unusual proximity effect to the
formation of the odd-frequency pairing state was not yet clarified.
Furthermore, there was no study of the proximity effect in junctions
with odd-frequency superconductors. The aim of the present paper is
to formulate a general theory of the proximity effect in the DN/S
junctions applicable to any type of symmetry state in a
superconductor forming the junction in the absence of spin-dependent
electronic scattering at the DN/S interface.
It will be shown that for spin-triplet [spin-singlet] superconductor junctions, odd-frequency spin-triplet even-parity (OTE) pairing state
[even-frequency spin-singlet even-parity (ESE) pairing state]
is generated in DN independent of the parity of the
superconductor. \par

Before proceeding with formal discussion, let us present qualitative
arguments illustrating the main conclusions of the paper. Two
constrains should be satisfied in the considered system: (1) only
the $s$-wave even-parity state is possible in the DN due to
isotropization by impurity scattering \cite{Sigrist}, (2) the spin structure of
induced Cooper pairs in the DN is the same as in an attached
superconductor. Then the Pauli principle provides the unique
relations between the pairing symmetry in a superconductor and the
resulting symmetry of the induced pairing state in the DN. Namely,
for even-parity superconductors, ESE and OTE states, the pairing symmetry
in the DN should remain ESE and OTE. On the other hand, for
odd-parity superconductors, even-frequency spin-triplet odd-parity (ETO)
and odd-frequency spin-singlet
odd-parity (OSO) states, the pairing symmetry in the
DN should be OTE and ESE, respectively.
The above results are based on general properties and
independent of the details of the geometry and the spin structure of the
spin-triplet superconductors. \par
The generation of the
OTE state in the DN attached to the ETO
p-wave superconductor  is of particular
interest. Similar OTE state can be generated in superconducting
junctions with diffusive ferromagnets
\cite{Efetov1,Efetov2} but due to
different physical mechanism. Although the symmetry properties can
be derived from the basic arguments given above, the quantitative
model has to be considered to prove the existence of nontrivial
solutions for the pair amplitude in the DN in each of the above
cases.

Let us start with the general symmetry properties of the quasiclassical
Green's functions in the considered system. The elements of retarded and
advanced Nambu matrices $\widehat{g}^{R,A}$
\begin{equation}
\widehat{g}^{R,A}=\left(
\begin{array}{cc}
g^{R,A} & f^{R,A} \\
\overline{f}^{R,A} & \overline{g}^{R,A}%
\end{array}%
\right)
\end{equation}%
are composed of the normal
$g_{\alpha ,\beta }^{R}(\bm{ r},\varepsilon ,%
\bm{p})$ and anomalous ${f}_{\alpha ,\beta }^{R}(\bm{
r},\varepsilon ,\bm{p})$ components with spin indices $\alpha$ and $\beta$.
Here $\bm{p}=\bm{p}_{F}/\mid \bm{p}%
_{F}\mid $, $\bm{p}_{F}$ is the Fermi momentum, $\bm{r}$ and $\varepsilon $
denote coordinate and energy of a quasiparticle measured from the Fermi
level.

The function $f^{R}$ and the conjugated function $\bar{f}^{R}$ satisfy the
following relation \cite{Serene,Eschrig}
\begin{equation}
\bar{f}_{\alpha ,\beta }^{R}(\bm{ r},\varepsilon ,\bm{p})=-[f_{\alpha ,\beta
}^{R}(\bm{ r},-\varepsilon ,\bm{-p})]^{\ast }.
\end{equation}

The Pauli principle is formulated in terms of the retarded and the advanced
Green's functions in the following way \cite{Serene}
\begin{equation}
f_{\alpha ,\beta }^{A}(\bm{ r},\varepsilon ,\bm{p})=-f_{\beta ,\alpha }^{R}(%
\bm{ r},-\varepsilon ,\bm{-p}).
\end{equation}%
By combining the two above equations, we obtain $\bar{f}_{\beta ,\alpha
}^{R}(\bm{ r},\varepsilon ,\bm{p})=[f_{\alpha ,\beta }^{A}(\bm{ r}%
,\varepsilon ,\bm{p})]^{\ast }$. Further, the definitions of the
even-frequency and the odd-frequency pairing are $f_{\alpha ,\beta }^{A}(%
\bm{ r},\varepsilon ,\bm{p})=f_{\alpha ,\beta }^{R}(\bm{ r},-\varepsilon ,%
\bm{p})$ and $f_{\alpha ,\beta }^{A}(\bm{ r},\varepsilon ,\bm{p})=-f_{\alpha
,\beta }^{R}(\bm{ r},-\varepsilon ,\bm{p})$, respectively. Finally we get
\begin{equation}
\bar{f}_{\beta ,\alpha }^{R}(\bm{ r},\varepsilon ,\bm{p})=[f_{\alpha ,\beta
}^{R}(\bm{ r},-\varepsilon ,\bm{p})]^{\ast }  \label{Even}
\end{equation}%
for the even-frequency pairing and
\begin{equation}
\bar{f}_{\beta ,\alpha }^{R}(\bm{ r},\varepsilon ,\bm{p})=-[f_{\alpha ,\beta
}^{R}(\bm{ r},-\varepsilon ,\bm{p})]^{\ast }  \label{Odd}
\end{equation}%
for the odd-frequency pairing. In the following, we will focus on
Cooper pairs with $S_{z}=0$ for the simplicity, remove the external
phase of the pair potential in the superconductor and  concentrate
on the retarded part of the Green's function. In the case of pairing
with $S_{z}=1$ our final results will not be changed. We consider a
junction consisting of a normal (N) and a superconducting reservoirs
connected by a quasi-one-dimensional diffusive conductor (DN) with a
length $L$ much larger than the mean free path. The interface
between the DN and the superconductor (S) at $x=L$ has a resistance
$R_{b}$ and the N/DN interface at $x=0$ has a resistance
$R_{b^{\prime }}$. For $R_{b^{\prime }}=\infty$, the present model
is reduced to the DN/S bilayer with vacuum at the DN free surface.
The Green's function in the superconductor can be parameterized as
$g_{\pm }(\varepsilon )\hat{\tau}_{3}+f_{\pm }(\varepsilon
)\hat{\tau}_{2}$ using Pauli matrices, where the suffix $+(-) $
denotes the right (left) going quasiparticles. $g_{\pm }(\varepsilon
)$ and $f_{\pm }(\varepsilon )$ are given by $g_{+}(\varepsilon
)=g_{\alpha ,\beta }^{R}(\bm{ r},\varepsilon ,\bm{p})$
$g_{-}(\varepsilon )=g_{\alpha ,\beta }^{R}(\bm{ r},\varepsilon
,\bar{\bm{p}})$ $f_{+}(\varepsilon
)=f_{\alpha ,\beta }^{R}(\bm{ r},\varepsilon ,\bm{p})$, and $%
f_{-}(\varepsilon )=f_{\alpha ,\beta }^{R}(\bm{ r},\varepsilon ,\bar{\bm{p}})
$, respectively, with $\bar{\bm{p}}=\bar{\bm{p}}_{F}/\mid \bm{p}_{F}\mid $
and $\bar{\bm{p}}_{F}=(-p_{Fx},p_{Fy})$. Using the relations (\ref{Even}), (%
\ref{Odd}), we obtain that $f_{\pm }(\varepsilon )=[f_{\pm }(-\varepsilon
)]^{\ast }$ for the even-frequency pairing and $f_{\pm }(\varepsilon
)=-[f_{\pm }(-\varepsilon )]^{\ast }$ for the odd-frequency pairing,
respectively, while $g_{\pm }(\varepsilon )=[g_{\pm }(-\varepsilon )]^{\ast }
$ in both cases.

In the DN region only the $s$-wave even-parity pairing state is allowed due
to isotropization by impurity scattering \cite{Sigrist}.
The resulting Green's function in
the DN can be parameterized by $\cos \theta \hat{\tau}_{3}+\sin \theta \hat{%
\tau}_{2}$ in a junction with an even-parity superconductor and by $\cos
\theta \hat{\tau}_{3}+\sin \theta \hat{\tau}_{1}$ in a junction with an
odd-parity superconductor. The function $\theta $ satisfies the Usadel
equation \cite{Usadel}
\begin{equation}
D\frac{\partial ^{2}\theta }{\partial x^{2}}+2i\varepsilon \sin \theta =0
\label{eq.1}
\end{equation}%
with the boundary condition at the DN/S interface \cite{Proximityd}
\begin{equation}
\frac{L}{R_{d}}(\frac{\partial \theta }{\partial x})\mid _{x=L}=\frac{%
\langle F_{1}\rangle }{R_{b}},  \label{eq.2}
\end{equation}%
\begin{equation}
F_{1}=\frac{2T_{1}(f_{S}\cos \theta _{L}-g_{S}\sin \theta _{L})}{%
2-T_{1}+T_{1}(\cos \theta _{L}g_{S}+\sin \theta _{L}f_{S})}
\end{equation}%
and at the N/DN interface
\begin{equation}
\frac{L}{R_{d}}(\frac{\partial \theta }{\partial x})\mid _{x=0}=\frac{%
\langle F_{2}\rangle }{R_{b^{\prime }}},\ \ F_{2}=\frac{2T_{2}\sin \theta
_{0}}{2-T_{2}+T_{2}\cos \theta _{0}},  \label{eq.3}
\end{equation}%
respectively, with $\theta _{L}=\theta \mid _{x=L}$ and $\theta _{0}=\theta
\mid _{x=0}$. Here, $R_{d}$ and $D$ are the resistance and the diffusion
constant in the DN, respectively. The brackets $\langle \ldots \rangle $
denote averaging over the injection angle $\phi $
\begin{equation}
\langle F_{1(2)}(\phi )\rangle =\int_{-\pi /2}^{\pi /2}d\phi \cos \phi
F_{1(2)}(\phi )/\int_{-\pi /2}^{\pi /2}d\phi T_{1(2)}\cos \phi ,
\label{average}
\end{equation}%
\begin{equation}
T_{1}=\frac{4\cos ^{2}\phi }{Z^{2}+4\cos ^{2}\phi },\;\;T_{2}=\frac{4\cos
^{2}\phi }{Z^{\prime }{}^{2}+4\cos ^{2}\phi },
\end{equation}%
where $T_{1,2}$ are the transmission probabilities, $Z$ and $Z^{\prime }$
are the barrier parameters for two interfaces. %
%
%
%
%
Here $g_{s}$ is given by $g_{S}=(g_{+}+g_{-})/(1+g_{+}g_{-}+f_{+}f_{-})$ and
$f_{S}=(f_{+}+f_{-})/(1+g_{+}g_{-}+f_{+}f_{-})$ for the even-parity pairing
and $f_{S}=i(f_{+}g_{-}-f_{-}g_{+})/(1+g_{+}g_{-}+f_{+}f_{-})$ for the
odd-parity pairing, respectively, with $g_{\pm }=\varepsilon /\sqrt{%
\varepsilon ^{2}-\Delta _{\pm }^{2}}$ and $f_{\pm }=\Delta _{\pm }/\sqrt{%
\Delta _{\pm }^{2}-\varepsilon ^{2}}$. $\Delta _{\pm }=\Delta \Psi (\phi
_{\pm })$ for even-frequency paring and $\Delta _{\pm }=\Delta
_{odd}(\varepsilon )\Psi (\phi _{\pm })$ for odd-frequency pairing, $\Psi
(\phi _{\pm })$ is the form factor with $\phi _{+}=\phi $ and $\phi _{-}=\pi
-\phi $.
$\Delta$ is the maximum value of the pair potential for even-frequency pairing.

In the following, we will consider four possible symmetry classes of
superconductor forming the junction and consistent with the Pauli principle:
ESE, ETO, OTE and OSO pairing states. We will use the fact that only
the even-parity $s$-wave pairing is possible in the DN due to the impurity
scattering and that the spin structure of pair amplitude in the DN is the
same\ as in an attached superconductor.

(1) Junction with ESE superconductor

In this case, $f_{\pm }(\varepsilon )=f_{\pm }^{\ast }(-\varepsilon )$ and $%
g_{\pm }(\varepsilon )=g_{\pm }^{\ast }(-\varepsilon )$ are satisfied. Then,
$f_{S}(-\varepsilon )=f_{S}^{\ast }(\varepsilon )=f_{S}^{\ast }$ and $%
g_{S}(-\varepsilon )=g_{S}^{\ast }(\varepsilon )=g_{S}^{\ast }$ and we
obtain for $F_{1}^{\ast }(-\varepsilon )$ 
\begin{equation*}
F_{1}^{\ast }(-\varepsilon )=\frac{2T_{1}[f_{S}\cos \theta _{L}^{\ast
}(-\varepsilon )-g_{S}\sin \theta _{L}^{\ast }(-\varepsilon )]}{%
2-T_{1}+T_{1}[\cos \theta _{L}^{\ast }(-\varepsilon )g_{S}+\sin \theta
_{L}^{\ast }(-\varepsilon )f_{S}]}.
\end{equation*}%
It follows from Eqs. \ref{eq.1}-\ref{eq.3} 
that 
$\sin \theta ^{\ast }(-\varepsilon )=\sin \theta (\varepsilon )$ and $%
\cos \theta ^{\ast }(-\varepsilon )=\cos \theta (\varepsilon )$.
Thus the
ESE state  is formed in the DN, in accordance with
the Pauli principle.

(2) Junction with ETO superconductor

Now we have $f_{\pm }(\varepsilon )=f_{\pm }^{\ast }(-\varepsilon )$ and $%
g_{\pm }(\varepsilon )=g_{\pm }^{\ast }(-\varepsilon )$. Then, $%
f_{S}(-\varepsilon )=-f_{S}^{\ast }(\varepsilon )=-f_{S}^{\ast }$ and $%
g_{S}(-\varepsilon )=g_{S}^{\ast }(\varepsilon )=g_{S}^{\ast }$. As a
result, $F_{1}^{\ast }(-\varepsilon )$ is given by
\begin{equation*}
F_{1}^{\ast }(-\varepsilon )=-\frac{2T_{1}[f_{S}\cos \theta _{L}^{\ast
}(-\varepsilon )+g_{S}\sin \theta _{L}^{\ast }(-\varepsilon )]}{%
2-T_{1}+T_{1}[\cos \theta _{L}^{\ast }(-\varepsilon )g_{S}-\sin \theta
_{L}^{\ast }(-\varepsilon )f_{S}]}.
\end{equation*}%
It follows from Eqs. \ref{eq.1}-\ref{eq.3} 
that $\sin \theta ^{\ast }(-\varepsilon )=-\sin \theta (\varepsilon )$ and 
$\cos \theta ^{\ast }(-\varepsilon )=\cos \theta (\varepsilon )$.
Thus the
OTE state is formed in the
DN. Remarkably, the appearance of the OTE state is the only
possibility to satisfy the Pauli principle, as we argued above.
Interestingly, the OTE pairing state can be also realized in
superconductor/ferromagnet junctions
\cite{Efetov1,Efetov2}, but the
physical mechanism differs from the one considered here.

(3) Junction with OTE superconductor

In this case $f_{\pm}(\varepsilon)=-f_{\pm}^{*}(-\varepsilon)$ and $
g_{\pm}(\varepsilon)=g_{\pm}^{*}(-\varepsilon)$. Then $f_{S}(-%
\varepsilon)=-f^{*}_{S}(\varepsilon)$ and $g_{S}(-
\varepsilon)=g^{*}_{S}(\varepsilon)$ and one can show that $%
F_{1}^{*}(-\varepsilon)$ has the same form as in the case of ETO
superconductor junctions. Then, we obtain $\sin
\theta^{*}(-\varepsilon)=-\sin\theta(\varepsilon)$ and $\cos
\theta^{*}(-\varepsilon)=\cos\theta(\varepsilon)$.
These relations mean that the OTE pairing state is
induced in the DN.

(4) Junction with OSO superconductor

We have $f_{\pm }(\varepsilon )=-f_{\pm }^{\ast }(-\varepsilon )$, $g_{\pm
}(\varepsilon )=g_{\pm }^{\ast }(-\varepsilon )$ and $f_{S}(-\varepsilon
)=f_{S}^{\ast }(\varepsilon )$, $g_{S}(-\varepsilon )=g_{S}^{\ast
}(\varepsilon )$. 
One can show that $F_{1}^{\ast }(-\varepsilon )$ takes the same form as in
the case of ESE superconductor junctions. 
Then, we obtain that $\sin \theta ^{\ast }(-\varepsilon )=\sin \theta
(\varepsilon )$ and $\cos \theta ^{\ast }(-\varepsilon )=\cos \theta
(\varepsilon )$. Following the same lines as in case (1), we conclude that
the ESE pairing state is induced in the DN.

We can now summarize the central conclusions in the table below.

\begin{center}
\begin{tabular}{|c|p{3cm}|p{3cm}|}
\hline
& Symmetry of the pairing in superconductors & Symmetry of the pairing in
the DN \\ \hline
(1) & Even-frequency spin-singlet even-parity (ESE) & ESE \\ \hline
(2) & Even-frequency spin-triplet odd-parity (ETO) & OTE \\ \hline
(3) & Odd-frequency spin-triplet even-parity (OTE) & OTE \\ \hline
(4) & Odd-frequency spin-singlet odd-parity (OSO) & ESE \\ \hline
\end{tabular}
\end{center}

Note that for even-parity
superconductors the resulting symmetry of the induced pairing state
in the DN is the same as that of a superconductor (the cases (1),
(3)). On the other hand, for odd-parity superconductors, the induced
pairing state in the DN has symmetry different from that of a
superconductor (the cases (2), (4)).

In order to illustrate the main features of the proximity effect in all the
above cases, we calculate the LDOS
$\rho(\varepsilon) = {\rm Real}[\cos\theta(\varepsilon)]$ and
the pair amplitude $f(\varepsilon)=\sin\theta(\varepsilon)$
in the middle of the DN layer at
$x=L/2$. We fix $Z=1$, $Z^{\prime}=1$, $R_{d}/R_{b}=1$,
$R_{d}/R_{b^{\prime}}=0.01$
and $E_{Th}=0.25\Delta$.

We start from junctions with ESE superconductors and choose the $s$-wave
pair potential with $\Psi _{\pm }=1$. The LDOS has a gap and the
Real(Imaginary) part of $f(\varepsilon )$ is an even(odd) function of $%
\varepsilon $ consistent with the formation of the even-frequency pairing.
\begin{figure}[tb]
\begin{center}
\scalebox{0.8}{
\includegraphics[width=7cm,clip]{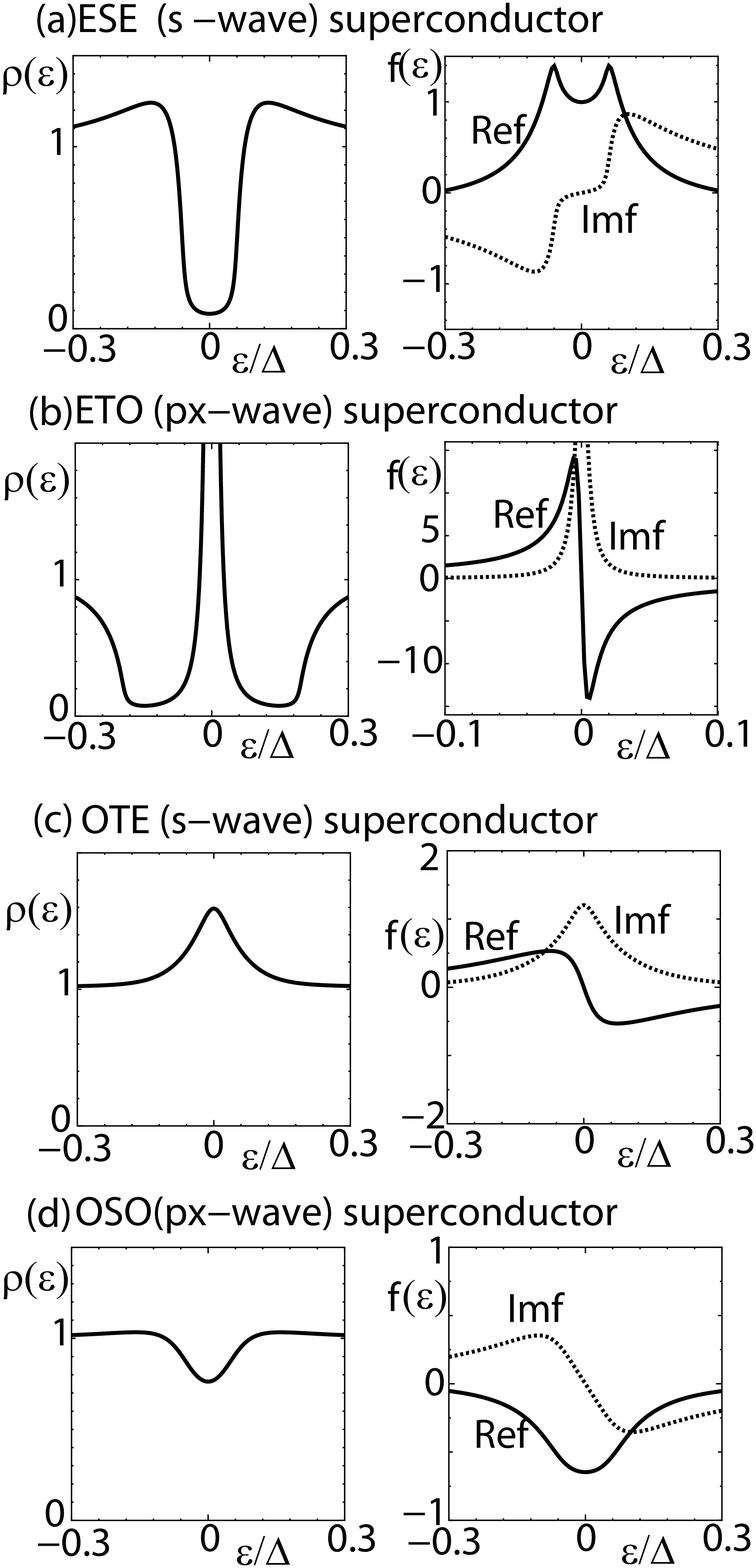}
}
\end{center}
\caption{Local density of states $\protect\rho (\protect\varepsilon )$ and
pair amplitude $f(\protect\varepsilon )$ at the center of the DN, $x=L/2$ is
plotted. Re$f$ and Im$f$ denote the  real and imaginary part of $f(%
\protect\varepsilon )$. The pairing symmetry of the superconductor is (a)
ESE, (b) ETO (c) OTE and (d) OSO, respectively. For (c) and (d), we choose $%
\bar{C}=0.8$. The resulting symmetry of $f(\protect\varepsilon )$ is
(a)ESE, (b) OTE (c) OTE and (d) ESE, respectively. }
\label{fig:1}
\end{figure}
In junctions with ETO superconductors, we choose $p_{x}$-wave pair potential
with $\Psi _{+}=-\Psi _{-}=\cos \phi $ as a typical example. In this case,
an unusual proximity effect is induced where the resulting LDOS has a zero
energy peak (ZEP) \cite{Proximityp}. The resulting LDOS has a ZEP \cite%
{Proximityp} since
$g^{2}(\varepsilon) + f^{2}(\varepsilon)=1$ and
$f(\varepsilon =0)$ becomes a purely imaginary
number. This is consistent with $f(\varepsilon )=-f^{\ast
}(-\varepsilon )$ and the formation of the OTE pairing in the DN.
To discuss junctions with an odd-frequency superconductor we choose
$\Delta ^{odd}(\varepsilon )=\bar{C}\varepsilon /[1+(\varepsilon
/\Delta)^{2}]$ as the simplest example of the $\varepsilon $
dependence of an odd-frequency superconductor pair potential.
At $\varepsilon=\Delta$, the magnitude of $\Delta ^{odd}(\varepsilon )$
becomes maximum.
Here we choose $\bar{C}<1$ when LDOS of bulk superconductor does not have
a gap around $\varepsilon =0$.
Let us first consider junctions with OTE superconductors and choose an $s$%
-wave pair potential as an example. The resulting LDOS has a ZEP, in
contrast to junctions with ESE superconductors where the resulting LDOS has
no ZEP. The formation of the ZEP is due to the similar reason in the
ETO superconductor junctions, where
$f(\varepsilon =0)$ is a pure imaginary number.
Finally, let us discuss junctions with OSO superconductors and choose $p_{x}$
-wave pair as an example. In this case, the ESE pairing is induced in the DN
and
$f(\varepsilon)=f^{*}(-\varepsilon)$ is satisfied. The resulting LDOS has a
gap since $f(\varepsilon=0)$ becomes a real number, in contrast to junctions
with OTE superconductors.

In summary, we have formulated a general theory of the proximity
effect in superconductor / diffusive normal metal junctions. Four
symmetry classes in a superconductor allowed by Pauli principle are
considered: 1) even-frequency spin-singlet even-parity (ESE), 2)
even-frequency spin-triplet odd-parity (ETO), 3) odd-frequency
spin-triplet even-parity (OTE) and 4) odd-frequency spin-singlet
odd-parity (OSO). We have found that the resulting symmetry of the
induced pairing state in the DN is 1) ESE 2) OTE 3) OTE and 4) ESE,
respectively. The symmetry in DN is established due to the
isotropization of the pair wave function by the impurity scattering
and spin conservation across the interface.
This universal feature is very important to classify
unconventional superconductors by
using proximity effect junctions.

One of the authors Y.T. expresses his sincerest gratitude to
clarifying discussions with M. Eschrig and Ya.V. Fominov.
Discussions with Y. Fuseya, K. Miyake, Yu. V. Nazarov, A.D.
Zaikin, A. F. Volkov and K. Efetov are gratefully acknowledged.
This work is supported by
Grant-in-Aid for Scientific Research
(Grant Nos. 17071007 and 17340106) from the Ministry of Education,
Culture, Sports, Science and Technology of Japan.


\end{document}